%% file: paper.tex
\documentclass[]{assets/fairmeta}

\usepackage{wrapfig}
\usepackage{tabularx}
\usepackage{textcomp}
\usepackage{stfloats}
\usepackage{url}
\usepackage{verbatim}
\usepackage{titlesec}
\usepackage{adjustbox}
\usepackage{multirow}
\usepackage{pifont}
\usepackage{tikz}
\usepackage{comment}
\usepackage{amsmath,amssymb} 
\usepackage{colortbl}  
\usepackage{color}
\usepackage{booktabs}
\usepackage{longtable}
\usepackage[numbers,sort&compress]{natbib}
\usepackage{hyperref}
\usepackage{graphicx}    
\usepackage{subcaption} 
\usepackage{multirow} 
\usepackage{booktabs} 
\usepackage{subcaption} 
\usepackage{array}
\usepackage{ragged2e}
\usepackage{makecell}
\usepackage{xeCJK}      
\RequirePackage{xspace}
\makeatletter
\DeclareRobustCommand\onedot{\futurelet\@let@token\@onedot}
\def\@onedot{\ifx\@let@token.\else.\null\fi\xspace}

\makeatother

\definecolor{adptorange}{RGB}{248, 205, 172}
\definecolor{cmpblue}{RGB}{189, 215, 238}
\definecolor{cmpblue}{RGB}{189, 215, 238}

\definecolor{our_red}{RGB}{232,157,160}
\definecolor{our_blue}{RGB}{136,206,230}
\definecolor{our_orange}{RGB}{246,200,168}
\definecolor{our_green}{RGB}{178,211,164}

\definecolor{attn_code0}{RGB}{247,215,200}
\definecolor{attn_code1}{RGB}{238,169,139}
\definecolor{mlp_code0}{RGB}{204,201,221}
\definecolor{mlp_code1}{RGB}{102,95,153}

\definecolor{token_blue}{RGB}{84, 120, 140}

\usepackage{pifont}       
\usepackage{bbding}       
\usepackage{fontawesome5}
\usepackage{xspace}

\usepackage{float}

\newlength\savewidth

\newcolumntype{x}[1]{>{\centering\arraybackslash}p{#1pt}}
\newcolumntype{y}[1]{>{\raggedright\arraybackslash}p{#1pt}}
\newcolumntype{z}[1]{>{\raggedleft\arraybackslash}p{#1pt}}

\renewcommand{\paragraph}[1]{\vspace{1mm}\noindent\textbf{#1}}
\usepackage{colortbl}
\usepackage{xcolor}
\usepackage{wrapfig}

\renewcommand{\paragraph}[1]{\vspace{1.25mm}\noindent\textbf{#1}}

\usepackage{algorithm}
\usepackage{listings}

\definecolor{codeblue}{rgb}{0.25, 0.5, 0.5}
\definecolor{codekw}{rgb}{0.35, 0.35, 0.75}
\lstdefinestyle{Pytorch}{
    language = Python,
    backgroundcolor = \color{white},
    basicstyle = \fontsize{9pt}{8pt}\selectfont\ttfamily\bfseries,
    columns = fullflexible,
    aboveskip=1pt,
    belowskip=1pt,
    breaklines = true,
    captionpos = b,
    commentstyle = \color{codeblue},
    keywordstyle = \color{codekw},
}

\definecolor{green}{HTML}{009000}
\definecolor{red}{HTML}{ea4335}

\title{DaX: Learning General Pathology Representations Across Scales}

\author[1,2]{Bokai Zhao}
\author[1]{Yiyang Zhang}
\author[1, 3]{Long Bai}
\author[1, 3]{Tai Ma}
\author[\dagger 1, 3]{Hanqing Chao}
\author[\dagger 1]{Minfeng Xu}

\affiliation[1]{DAMO Academy, Alibaba Group\\}
\affiliation[2]{Institute of Automation, Chinese Academy of Sciences} 
\affiliation[3]{Hupan Lab}
\contribution[\dagger]{Corresponding authors}

\input{sec/0_abstract}

\date{\today} 

\begin{document}

\thispagestyle{firstheader}
\maketitle
\pagestyle{empty}

\input{sec/1_introduction}
\input{sec/2_method}
\input{sec/3_results}
\input{sec/4_discussion}

\bibliographystyle{unsrtnat}
\bibliography{reference/pathology_reference,reference/dataset,reference/general_method,reference/Pathology_model}

\newpage
\beginappendix
\input{sec/appendix}

\end{document}

%% file: sec/0_abstract.tex
\abstract{

Computational pathology requires visual representations that generalize across diverse clinical endpoints, including diagnosis, grading, staging, biomarker prediction, treatment response, prognosis, and tissue source recognition. These tasks rely on morphologic evidence at multiple scales, from nuclei-level detail to large-scale tissue architecture, and are affected by variation in magnification, staining, scanner type, slide preparation, and input resolution. Existing pathology foundation models have substantially advanced the field, but continuous scale modeling, dense feature stability, input-size flexibility, and statistically rigorous evaluation are often addressed separately. In this technical report, we present \textbf{DaX (大象)}, a pathology vision foundation model designed to integrate these requirements within a unified framework. DaX follows a DINOv3-style self-supervised training framework initialized from natural-image DINOv3 weights and adapts general multi-scale visual priors to histopathology through continuous magnification training, redesigned cross-scale self-supervision, pathology-specific augmentation, multi-input-size training, and Gram-anchored consistency. We further construct a broad benchmark of whole-slide prediction tasks, including 161 tasks from 44 public datasets, covering 28,182 patients and 34,394 slides across 4 primary task categories and 9 subcategories. All models are evaluated under a fixed cross-fold protocol with patient-disjoint splits and statistical ranking based on fold-level significance tests, enabling reproducible and statistically grounded comparisons that are less sensitive to split-dependent variation. Under this rigorous setting, DaX shows strong and stable transfer performance across diagnostic, molecular, prognostic, and specimen-context tasks. These results support DaX as a general visual encoder for pathology and provide a reproducible evaluation framework for future pathology foundation models. The project page can be accessed at \url{https://alibaba-damo-academy.github.io/DaX/benchboard/}.
}

%% file: sec/1_introduction.tex
\section{Introduction} \label{sec:introduction}
\noindent 

Digital pathology has become increasingly central to modern cancer diagnosis, biomarker discovery, and computational pathology research~\citep{Digital-pathology-and-artificial-intelligence}. In routine practice, pathologists interpret whole-slide images (WSIs) by moving across multiple magnifications. Low-power views provide information about tissue architecture, invasion patterns, and glandular organization, whereas high-power views reveal cellular morphology, nuclear atypia, and mitotic activity. This multi-resolution workflow makes pathology image understanding inherently hierarchical and strongly scale-dependent~\citep{Artificial-intelligence-for-digital-and-computational-pathology}. An effective pathology foundation model should therefore capture both global tissue-level context and fine-grained local morphology. It should also remain robust to variations in staining, slide preparation, scanner characteristics, and data source. These properties are essential for broad downstream use, including classification, segmentation, risk modeling, biomarker prediction, and multimodal pathology systems~\citep{A-survey-of-pathology-foundation-model:-Progress-and-future-directions}.

Recent pathology foundation models have substantially advanced representation learning for histopathology. Prior work has explored large-scale self-supervised pretraining, vision-language alignment, whole-slide modeling, and clinical-scale representation learning. Representative models include pathology vision encoders such as UNI~\citep{UNI}, Prov-GigaPath~\citep{Prov-GigaPath}, Virchow~\citep{Virchow}, and Phikon-v2~\citep{Phikon-v2}; vision-language models such as PLIP~\citep{PLIP}, CONCH~\citep{CONCH} and MUSK~\citep{MUSK}; the multimodal model mSTAR~\citep{mSTAR}; and the expert-distillation model GPFM~\citep{GPFM}. These models have improved the quality and transferability of pathology features. Nevertheless, several practical requirements remain insufficiently integrated into a unified pretraining objective. First, many models are trained at fixed or limited magnifications, despite the fact that pathology semantics change continuously across scale. Second, local-global crop strategies are often inherited from natural-image self-supervision and may not reflect the biological relationship between nuclei-level detail and tissue-level organization~\citep{Dinov2}. Third, many encoders are trained under a relatively fixed input size, which limits their flexibility across downstream tasks with different fields of view and memory constraints~\citep{Mind-the-Gap}. In parallel, evaluation protocols often rely on limited datasets, task types, or data splits, making it difficult to assess broad generalization and the statistical reliability of model comparisons~\citep{Benchmarking-foundation-models-as-feature-extractors-for-weakly-supervised-computational-pathology}.

These limitations point to two coupled challenges in pathology foundation modeling: how to learn representations that reflect the multi-scale nature of pathology, and how to evaluate such representations under realistic downstream conditions. A general pathology encoder should operate across a wide range of magnifications, associate local cellular evidence with broader tissue context, remain stable under stain and scanner variation, support different input sizes, and preserve dense local features. At the same time, its evaluation should reflect the diversity of real pathology applications, which span diagnosis, grading, staging, molecular alteration, immune phenotype, treatment response, survival, and tissue origin. Narrow benchmarks may overestimate generalization, and single-split evaluations can make performance sensitive to data partitioning. A stronger evaluation protocol should therefore cover diverse clinical endpoints, use fixed patient-level splits, and report improvements supported by statistical evidence across folds.

In this work, we present \textbf{DaX (大象)}, a pathology vision foundation model designed for general multi-scale pathology representation learning. DaX uses a ViT-L backbone and follows a DINOv3-style self-supervised training framework. It is initialized from natural-image DINOv3 weights~\citep{Dinov3}, allowing it to inherit general multi-scale visual priors before pathology-specific adaptation. In Stage 1, DaX is adapted to histopathology through continuous magnification training, pathology-specific augmentation, and redesigned cross-scale self-supervision. By enlarging the scale gap between local and global views, this stage encourages student-teacher alignment to connect nuclei-level morphology with broader tissue organization. In Stage 2, the representation is further refined through multi-input-size training and Gram-anchored consistency, improving robustness to input resolution while stabilizing dense feature geometry.

For evaluation, we construct a WSI-level benchmark containing 161 tasks from 44 public datasets (Table~\ref{tab_1}), covering 28,182 patients and 34,394 slides. The benchmark spans 4 primary task categories and 9 subcategories, covering diagnostic, molecular, prognostic, and specimen-context endpoints(Table~\ref{tab_1a}). All models are evaluated under a fixed patient-level cross-fold protocol with a statistical ranking strategy based on pairwise significance tests over fold-level results. Across this broad evaluation setting, DaX demonstrates strong and stable transfer performance, supporting its potential as a robust visual backbone for clinically diverse pathology applications.

The main contributions of this technical report are as follows:
\begin{itemize}
    \item We introduce \textbf{DaX (大象)}, a pathology-specific foundation model framework that leverages natural-image DINOv3 initialization as a source of general multi-scale visual priors and adapts it to histopathology through unified self-supervised pretraining, explicitly targeting continuous magnification modeling, stain and acquisition robustness, input-size flexibility, and pathology-aware local-global alignment.
    
    \item We develop a two-stage training strategy that integrates continuous magnification modeling, stain and acquisition robustness, input-size flexibility, pathology-aware local-global alignment, and dense feature stabilization into a unified pretraining framework.
    
    \item We establish a broad benchmark of whole-slide prediction tasks, including 161 tasks from 44 public datasets and organized into 4 primary task categories and 9 subcategories covering diagnostic, molecular, prognostic, and specimen-context endpoints.
    
    \item We provide a fixed cross-fold evaluation protocol with patient-disjoint splits and statistical ranking, enabling more reliable model comparison than single-split evaluation and showing that DaX achieves strong transfer performance across the full benchmark.
\end{itemize}

%% file: sec/2_method.tex
\section{Methodology}
\noindent

\subsection{Data Collection and Benchmark Construction}

\subsubsection{Pretraining Data Collection}

The objective of DaX is to learn pathology representations that remain robust across magnification, input size, tissue type, and data source. To support this goal, we construct a large-scale pretraining data collection comprising 104,569 WSIs from three public resources: TCGA~\citep{TCGA}, GTEx~\citep{GTEx}, and HistAI~\citep{Histai}. These sources provide complementary coverage of normal, precancerous, and malignant tissue morphology. As summarized in Figure~\ref{fig_1}(A), the resulting collection spans a broad spectrum of organ systems and is characterized at both the patient and slide levels. The slide-level distribution reflects the scale of available training images, whereas the patient-level distribution demonstrates cohort diversity. Together, these statistics indicate substantial morphological variation across organs, diseases, and public data sources, thereby mitigating potential biases toward any single tissue type, scanner condition, or source dataset during pretraining~\citep{Towards-robust-foundation-models-for-digital-pathology}.

We apply a unified preprocessing pipeline to all WSIs prior to pretraining. Each WSI is resampled to a common physical resolution of $0.5~\mu\mathrm{m/pixel}$ at $20\times$ magnification to establish a shared spatial scale. Tissue foreground regions are then segmented, and empty background areas are discarded. From the retained regions, we extract $1920 \times 1920$-pixel patches at four anchor magnifications: $2.5\times$, $5\times$, $10\times$, and $20\times$. Adjacent patches are extracted with a 640-pixel overlap, and patches with a tissue foreground ratio below two-thirds are filtered out. This procedure yields a calibrated, multi-resolution patch pool that captures both low-magnification tissue architecture and high-magnification cellular morphology, serving as the data foundation for the continuous magnification training described below.

\subsubsection{Benchmark Dataset and Task Construction}

\begin{figure}
\centering
\captionsetup{
  labelfont={bf,normalsize},
  font=normalsize,
  skip=8pt
}
\includegraphics[width=0.95\linewidth]{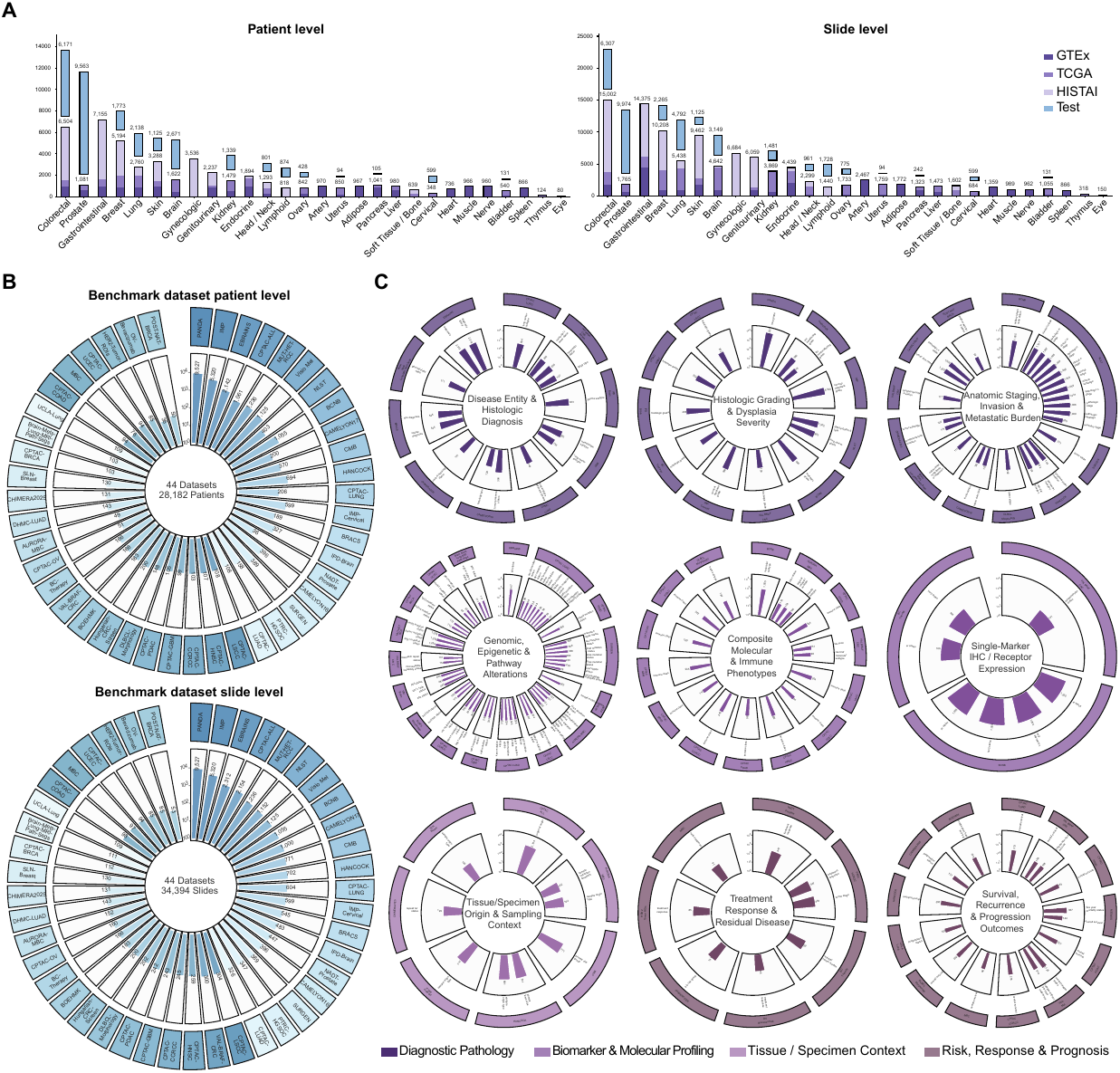}
\caption{
\textbf{Data composition for DaX pretraining and benchmark evaluation.}\vspace{0.1cm} \newline 
\small
(A) Distribution of sample sizes across tissue sources. The left and right panels summarize the number of patients and WSIs, respectively. Bars are grouped by tissue or organ type and color-coded by data source (GTEx, TCGA, HISTAI, and independent test datasets).
(B) Composition of the benchmark datasets. Circular bar plots detail the sample distribution across 44 benchmark datasets (28,182 patients; 34,394 WSIs). Each sector represents an individual dataset, with the radial bar height denoting the corresponding number of patients or WSIs, and outer labels indicating dataset names.
(C) Task-level organization of the benchmark. Downstream tasks are categorized into four color-coded major groups: (1) diagnostic pathology; (2) biomarker and molecular profiling; (3) tissue/specimen context; and (4) risk, response, and prognosis. Across the nine circular bar plots, each sector denotes a specific task, with the radial bar height representing the WSI count and the outer segments indicating the contributing datasets.}

\label{fig_1}
\end{figure}

To systematically evaluate pathology foundation models, we curated a WSI-level benchmark from 44 independent public pathology datasets with available clinical, pathological, molecular, or outcome annotations. Comprising 28,182 patients and 34,394 WSIs (Figure~\ref{fig_1}(B)), this benchmark spans diverse organs, disease entities, molecular endpoints, clinical outcomes, and specimen contexts. The benchmark datasets are strictly disjoint from the pretraining data collection, ensuring that downstream evaluations assess held-out generalization. Rather than treating each dataset as a single evaluation task, we systematically extracted clinically meaningful prediction endpoints from the available metadata and converted them into standardized benchmark tasks.

Through this process, we defined 161 benchmark tasks. Each task is specified by four components: a source dataset, an eligible patient or WSI cohort, a prediction target, and a task type. We retained labels only when they corresponded to clinically relevant endpoints, such as histologic grading, molecular alterations, treatment response, or survival outcomes. To improve label consistency, we harmonized annotations within each dataset by removing invalid or ambiguous entries, merging clinically equivalent labels when appropriate, and filtering out tasks with insufficient sample sizes or unstable label definitions. For datasets containing multiple WSIs per patient, cross-validation folds were strictly partitioned at the patient level to preclude data leakage across the training, validation, and test splits.

To facilitate structured evaluation, we organized the 161 benchmark tasks into a two-level clinical taxonomy. At the coarse level, tasks are grouped into four clinical domains: diagnostic pathology; biomarker and molecular profiling; tissue/specimen context; and risk, response, and prognosis. These domains are further divided into nine fine-grained task categories, such as histologic grading and dysplasia severity, composite molecular and immune phenotypes, and treatment response and residual disease. Figure~\ref{fig_1}(C) summarizes the WSI distribution across these task categories, while Table ~\ref{tab_1} provides the complete task inventory and taxonomy mapping. This hierarchical design enables both global model comparison and granular performance analysis across clinically distinct task groups.

\subsection{DaX Foundation Model}

DaX is built upon a ViT-L backbone initialized with DINOv3 weights and optimized using a self-supervised student--teacher framework~\citep{Dinov3}. During pretraining, the teacher network is updated by exponential moving average to provide stable targets for augmented student views. Following a DINO/iBOT-style objective, DaX applies self-supervised alignment to both class tokens and patch tokens, enabling the model to learn global tissue-level semantics while preserving local morphology-aware token representations.

Although this initialization and training framework provide a strong general visual prior, WSIs introduce domain-specific requirements that are not fully addressed by natural-image pretraining. Histopathology interpretation depends on hierarchical visual evidence across magnifications, ranging from low-magnification tissue architecture to intermediate glandular organization and high-magnification cellular morphology. Pathology images also lack a fixed canonical orientation and exhibit substantial stain, scanner, and preparation-related heterogeneity across data sources. In addition, many downstream tasks rely on dense local features that capture the intrinsic spatial organization of tissue. These dense representations can be sensitive to input resolution and may undergo geometric instability during extended self-supervised training. These considerations motivate a pathology-specific pretraining strategy that jointly addresses the magnification continuum, appearance shifts, local--global spatial correspondence, and dense feature stability.

To this end, DaX adopts the two-stage training framework shown in Figure~\ref{fig_2}(A). Stage 1 focuses on pathology-specific representation learning by integrating continuous magnification sampling, cross-scale multi-view generation, and orientation-agnostic and acquisition-robust augmentation. This stage is designed to link localized cellular morphology with broader tissue architecture while improving robustness to stain, focus, and acquisition-related variation. Stage 2 performs scale-aware dense refinement. Starting from the converged Stage 1 model, this stage introduces multi-size crop training and Gram-anchored dense consistency to improve compatibility across variable input resolutions and stabilize token-level feature geometry. Together, these two stages enable DaX to learn transferable pathology representations across magnification shifts, staining heterogeneity, and diverse downstream inference settings.

\subsubsection{Stage 1: Pathology-Specific Representation Learning}

Stage 1 adapts the DINOv3-initialized ViT-L backbone to pathology-specific visual structures. It focuses on three components: pathology adaptation from natural-image initialization, continuous magnification learning with cross-scale tissue views, and orientation-agnostic, acquisition-robust morphology learning. These designs aim to connect local cellular morphology with broader tissue-level context.

\textbf{Pathology adaptation from DINOv3 initialization.}
DaX is initialized with DINOv3 weights pretrained on large-scale natural images. This initialization provides a strong visual prior and improves optimization stability compared with training from scratch~\citep{kaiko}. Although histopathology images differ from natural images in texture, color distribution, and spatial organization, generic visual priors such as shape grouping, structural continuity, and part--whole relations remain useful. We therefore use DINOv3 initialization as the starting point and adapt the model through pathology-specific self-supervised pretraining.

\textbf{Continuous magnification learning with cross-scale tissue views.}
Pathology interpretation relies on visual evidence across magnifications. Low-magnification views capture tissue architecture and regional organization, whereas high-magnification views reveal cellular morphology, nuclear atypia, mitotic activity, and other fine-grained cues. Training at a fixed magnification, or only at several discrete magnifications, may limit representation continuity across scales. Prior work also suggested that discrete magnification training can weaken generalization at intermediate scales~\citep{Mind-the-Gap}.
DaX is therefore trained with continuous magnification variation over the $2.5\times$--$20\times$ range. Starting from the multi-resolution patch pool, we sample patches from four anchor magnifications and generate training views through random cropping and scaling. This approximates a continuous magnification spectrum and reduces dependence on fixed acquisition or pyramid levels.

We also increase the scale separation between local and global views. Local views are sampled from smaller regions and emphasize fine morphology, such as nuclei, glandular fragments, and stromal details. Global views cover larger tissue regions and provide architectural context. Because these views are generated from the same anchor region, they remain spatially related while presenting different levels of visual evidence. Alignment across these views is designed to link localized cellular morphology with broader tissue architecture.

\textbf{Orientation-agnostic and acquisition-robust morphology learning.}
Histopathology images have no fixed canonical orientation, and tissue rotation usually does not change diagnostic semantics. We therefore apply arbitrary-angle rotations to mitigate orientation bias and encourage morphology-centered representations. This design is consistent with prior evidence that rotation-agnostic pathology pretraining improves transfer across datasets~\citep{Rotation-agnostic}.

We further apply acquisition-oriented perturbations to improve robustness across centers, scanners, staining protocols, and slide preparation procedures. These include brightness, contrast, saturation, and hue jittering to simulate stain and illumination variation, together with Gaussian blur to mimic out-of-focus acquisition and local imaging degradation. By exposing the student network to diverse color, focus, and acquisition conditions, DaX is encouraged to rely more on tissue morphology and spatial organization than on unstable acquisition-specific cues.

\subsubsection{Stage 2: Scale-Aware Dense Refinement}

Stage 2 further refines the converged Stage 1 model to support variable input sizes and improve dense local representation quality. While Stage 1 establishes continuous magnification adaptation and local--global pathology correspondence, Stage 2 addresses a practical deployment requirement: the same foundation model should remain stable across different crop sizes, fields of view, and inference settings. This is important in digital pathology, where the same tissue pattern may occupy different spatial support due to patch extraction protocols, scanner resolutions, or hardware memory constraints. Stage 2 is therefore designed to improve input-size compatibility and stabilize token-level feature geometry while preserving the global semantic structure learned in Stage 1.

\begin{figure}
\centering
\captionsetup{
  labelfont={bf,normalsize},
  font=normalsize,
  skip=4pt
}
\includegraphics[width = \linewidth ]{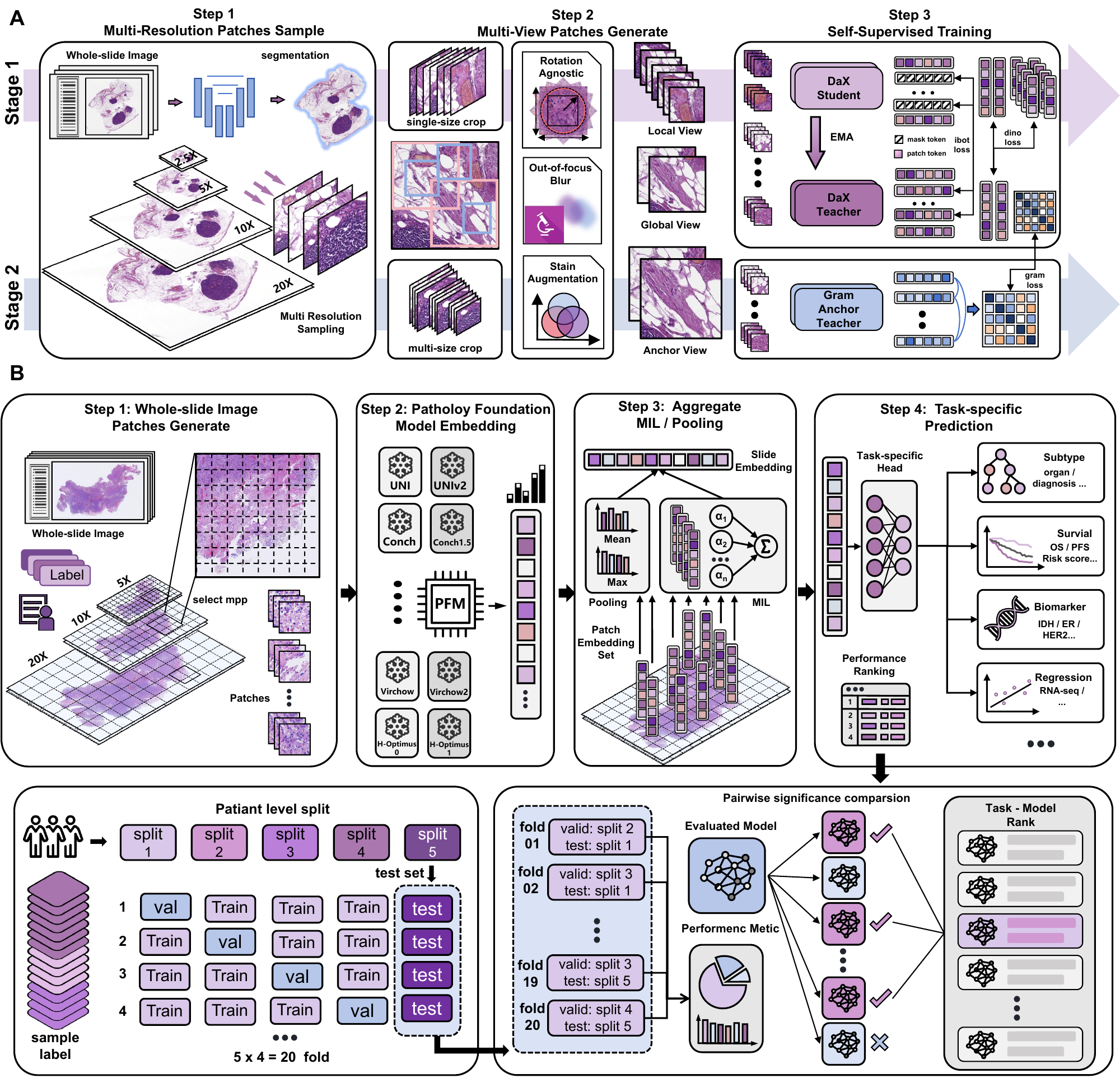}
\caption{
\textbf{Training framework and benchmark evaluation protocol of DaX.}
\vspace{0.1cm} \newline
\small
(A) Pathology-specific pretraining framework. DaX uses a two-stage DINOv3-style self-supervised framework. Stage 1 captures hierarchical tissue context through multi-resolution sampling and pathology-specific augmentations to generate multi-view inputs. The student and teacher networks are optimized using global representation alignment, patch-token alignment, and masked patch-token objectives. Stage 2 applies multi-size crops and Gram-anchored consistency for scale-aware dense refinement, stabilizing local representations across variable input sizes.
(B) Standardized evaluation protocol. Tissue foregrounds are extracted from WSIs, tiled into patches, and processed by a frozen encoder. Patch embeddings are aggregated using mean pooling or ABMIL for downstream prediction. Evaluation is performed using a structured $5 \times 4$ patient-level cross-validation design, yielding 20 folds. Final model rankings are based on pairwise significance testing of a predefined major metric, ensuring robust and statistically supported comparisons.
}
\label{fig_2}
\end{figure}

\textbf{Multi-input-size adaptation.}
We first introduce multi-size crop training, as shown in Figure~\ref{fig_2}(A). Specifically, we use three global--local crop pairs: $(512, 224)$, $(384, 168)$, and $(768, 336)$. These settings cover smaller, medium, and larger input resolutions, and all views are processed by the same backbone with shared parameters. This strategy exposes the encoder to different effective receptive fields during training and encourages a unified representation space across input sizes, rather than resolution-specific behavior.

\textbf{Gram-anchored dense consistency.}
To further stabilize dense feature geometry, we introduce Gram-anchored consistency during Stage 2. Dense pathology tasks often depend on local coherence, boundary separability, and the spatial organization of neighboring tissue patterns. Instead of directly matching token feature values, we regularize the pairwise similarity structure of token-level features. Specifically, the Gram matrix captures token-to-token similarity and is anchored to a stable teacher representation from an earlier training stage. This provides an additional refinement signal that constrains dense feature geometry while allowing sufficient flexibility in feature values. In Stage 2, the model is trained with a $2\times$ global-resolution setting, and the largest global view reaches $1536 \times 1536$ pixels. This high-resolution setting exposes the model to a larger spatial field and increases the demand for local feature coherence.

\subsection{Benchmark Evaluation Protocol}

We employ a standardized and reproducible evaluation protocol across all benchmark tasks, as illustrated in Figure~\ref{fig_2}(B). The protocol consists of four main steps: patient-level fold construction, slide-level feature extraction, task-specific prediction, and statistical model ranking. All compared models are evaluated under identical data splits, patch-processing configurations, aggregation strategies, and task-specific metrics.

\textbf{Evaluated foundation models.}
We evaluate DaX alongside a comprehensive set of existing foundation models, spanning pathology vision models, vision--language models, expert-distilled models, multimodal and spatial-transcriptomics-aligned models, and natural-image baselines. For each model, we document the model type, pretraining method, pretraining magnification and spatial resolution, patch encoder architecture, parameter count, and reported pretraining data scale. These attributes, summarized in Table~\ref{tab_2}, provide the model-level context for subsequent scale--performance analyses and task-level ranking results.

\textbf{Patient-level fold construction.}
For each benchmark task, labeled samples are partitioned into five patient-level splits. One split is reserved as the test set, while the remaining four splits form the non-test set. Within the non-test set, one split is iteratively selected as the validation set, with the remaining three used for training. Repeating this process over all five test-split choices yields a structured $5 \times 4$ patient-level cross-validation design with 20 folds. The same folds are used for all compared models to ensure deterministic evaluation and avoid patient-level data leakage.

\textbf{Slide-level feature extraction.}
For each WSI, tissue foreground is first delineated. The retained tissue regions are then tiled into patches at a fixed predefined magnification, either $10\times$ or $20\times$. These patches are processed by a frozen foundation model encoder to extract patch-level embeddings, converting each WSI into a bag of feature vectors.

\textbf{Feature aggregation and prediction.}
The extracted patch embeddings are aggregated into a slide-level representation using either mean pooling or attention-based multiple instance learning (ABMIL~\citep{ABMIL}). A task-specific prediction head is then trained on top of this slide-level representation. This setting evaluates the transferability of frozen foundation model features under standardized downstream readouts.

\textbf{Task evaluation.}
Each fold produces predictions on its held-out test split. These predictions are evaluated using the predefined major metric for the corresponding task. The major metric is selected according to task type, including classification metrics for categorical endpoints, survival-specific metrics for time-to-event endpoints, and continuous-outcome metrics for regression tasks. Consequently, each model--task--aggregation combination yields 20 fold-level test scores under a unified evaluation measure.

\textbf{Statistical ranking.}
For each task, model comparison is based on fold-level statistical ranking rather than raw average performance alone. We perform pairwise significance testing between each evaluated model and every competing model using the 20 fold-level scores of the predefined major metric. The task-level ranking score $R_{m,t}$ for model $m$ on task $t$ is defined as the number of competing models that are significantly outperformed by model $m$:
\begin{equation}
R_{m,t} =
\sum_{j \neq m}
\mathbb{I}
\left[
p_{m,j,t} < \alpha
\land
\bar{s}_{m,t} > \bar{s}_{j,t}
\right],
\end{equation}
where $\bar{s}_{m,t}$ denotes the average test score across 20 folds, $p_{m,j,t}$ is the $p$-value from the paired significance test between models $m$ and $j$ on task $t$, $\alpha$ is the significance threshold, and $\mathbb{I}[\cdot]$ is the indicator function. This ranking rule is applied consistently across all tasks and aggregation strategies, providing the basis for the task-level heatmaps in Figure~\ref{fig_3}(D).

%% file: sec/3_results.tex
\section{Experimental Results}

We evaluated DaX against a broad collection of pathology foundation models, vision--language models, multimodal models, spatial-transcriptomics-aligned models, and natural-image baselines under the standardized benchmark protocol. The results are analyzed at three levels: overall benchmark performance and scaling trends, performance across task categories, and task-level statistical rankings across all 161 benchmark tasks.

\subsection{Overall Benchmark Performance and Scaling Trends}

Figure~\ref{fig_3}(A) summarizes the relationship between pretraining data scale, model capacity, and overall benchmark performance. Each bubble represents a foundation model: the x-axis denotes the reported number of training WSIs on a logarithmic scale, the y-axis represents the mean performance across benchmark tasks, and the bubble size indicates the parameter count. DaX-Base and DaX represent two model scales within our pretraining framework, using ViT-B and ViT-L backbones, respectively. This paired configuration allows us to examine how the proposed pathology-specific pretraining strategy scales with model capacity.

DaX achieves the highest mean performance across benchmark tasks among all evaluated models. Compared with DaX-Base, the ViT-L variant further improves mean performance, indicating a positive scaling trend within the proposed framework. More broadly, the distribution in Figure~\ref{fig_3}(A) shows a clear scale--performance relationship: models pretrained on larger WSI collections generally achieve stronger average performance, although substantial dispersion remains across models. This trend suggests that large-scale pathology pretraining is an important driver for learning transferable WSI representations.

However, pretraining scale and parameter count alone do not fully determine downstream performance. Models with comparable training data sizes or parameter counts show notable variation in benchmark performance, indicating that pretraining strategy, magnification modeling, input-size adaptation, dataset composition, and pathology-specific augmentation also contribute substantially to model transferability. The leading average performance of DaX suggests that the two-stage training design---combining continuous magnification learning with scale-aware dense refinement---enhances the general utility of representations across heterogeneous WSI-level tasks.

Prior benchmark studies have shown that scaling effects in pathology foundation models can depend strongly on the evaluation setting: model-size effects were weak for detection tasks and more evident but task-dependent for biomarker prediction tasks, while pretraining dataset size showed no strong correlation with downstream performance~\citep{clinicalbenchmark}. Our results provide a complementary observation: under a statistically aggregated evaluation across a broader set of WSI-level tasks, the reported training WSI scale exhibits a clearer positive relationship with overall mean performance. At the same time, the performance variation among similarly scaled models further highlights that methodological design and dataset composition remain critical beyond scale alone.

\subsection{Performance across Task Categories}

Figure~\ref{fig_3}(B) compares representative models across nine fine-grained task categories, while Figure~\ref{fig_3}(C) summarizes task-wise performance distributions across four coarse-grained clinical domains. These analyses reveal that model performance is highly task-dependent. Diagnostic pathology tasks generally exhibit higher and more stable performance, reflecting their direct dependence on H\&E morphology and histologic patterns. In contrast, biomarker and molecular profiling tasks, as well as risk, response, and prognosis-related tasks, show larger variability, consistent with the weaker and more indirect association between morphology and molecular or outcome labels.

Across the nine fine-grained categories, DaX achieves consistently strong performance and exhibits a balanced profile among the evaluated models. This indicates that the proposed pretraining strategy is not limited to morphology-dominant diagnostic tasks, but also transfers to more challenging endpoints such as molecular alterations, immune phenotypes, treatment responses, and prognosis. The radar plot further suggests that DaX maintains strong category-level performance without substantial performance drops in categories where many baseline models show weaker transfer.

At the coarse-grained domain level, the box plots in Figure~\ref{fig_3}(C) provide a task-wise view of performance dispersion. DaX shows high median performance and stable distributions across diagnostic pathology, biomarker and molecular profiling, tissue/specimen context, and risk, response, and prognosis. This pattern suggests that the performance of DaX is not driven by a small number of outlier tasks, but is distributed across clinically distinct task groups. Together, the fine-grained and coarse-grained analyses support the broad transferability of DaX across heterogeneous WSI-level prediction tasks.

\subsection{Task-Level Statistical Ranking across Benchmark Tasks}

\begin{figure}
\centering
\captionsetup{
  labelfont={bf,normalsize},
  font=normalsize,
  skip=4pt
}
\includegraphics[width = \linewidth ]{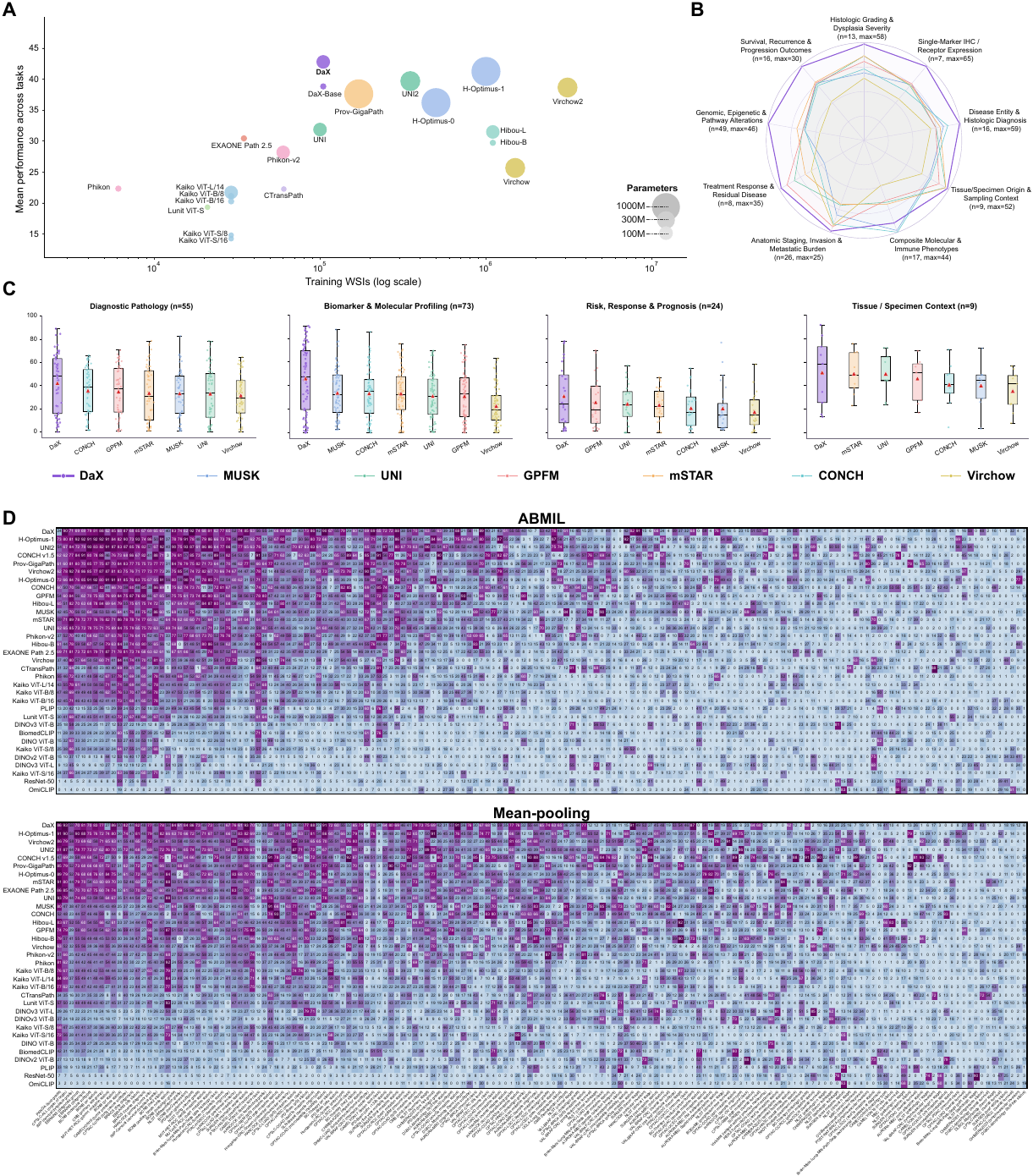}
\caption{
\textbf{Benchmark results and model ranking scores across pathology foundation models.}\vspace{0.1cm} \newline
\small
(A) Relationship between pretraining scale, model size, and overall performance. Each bubble represents a foundation model. The x-axis denotes pretraining WSIs (log scale), the y-axis indicates mean performance across all tasks, and bubble size reflects parameter count.
(B) Fine-grained category-level performance comparison. The radar plot summarizes the mean performance of representative models across nine specific task categories.
(C) Coarse-grained domain-level performance distributions. Box plots display task-wise performance aggregated into four higher-level clinical domains. Points represent individual task scores, while boxes summarize intra-domain distributions.
(D) Comprehensive task-level ranking heatmaps. Heatmaps report model rankings across all 161 benchmark tasks under ABMIL (top) and mean-pooling (bottom) aggregations. Rows correspond to evaluated models and columns to tasks. Cell values indicate task-specific ranking scores from pairwise significance testing, where higher values denote significant superiority over a greater number of competing models.
}
\label{fig_3}
\end{figure}

Figure~\ref{fig_3}(D) reports task-level ranking heatmaps across all benchmark tasks under both ABMIL and mean-pooling aggregation strategies. Unlike raw average performance, the ranking score is derived from pairwise significance testing over fold-level results and reflects the number of competing models that are significantly outperformed on each task. This design provides a statistically grounded comparison, reducing the influence of small numerical differences and split-specific variability.

DaX obtains high ranking scores across a broad set of tasks under both aggregation strategies. The consistency between ABMIL and mean pooling indicates that the strong performance of DaX is primarily driven by the quality of the learned patch-level representations, rather than by a specific slide-level aggregation method. In comparison, several existing models show strong performance only in selected task subsets or under a single aggregation strategy, suggesting less stable transfer across heterogeneous clinical endpoints.

Overall, the task-level ranking analysis shows that DaX provides broadly transferable pathology representations and achieves statistically supported advantages across diverse WSI-level benchmark tasks.

%% file: sec/4_discussion.tex
\section{Discussion and Conclusion}
\noindent

This report presents DaX, a pathology-specific visual foundation model designed for multi-scale and spatially organized histopathology representation learning. DaX combines continuous magnification learning, cross-scale tissue views, orientation-agnostic and acquisition-robust augmentation, multi-input-size adaptation, and Gram-anchored dense refinement. These components adapt a DINOv3-initialized ViT backbone to the practical requirements of computational pathology, where diagnostic evidence may appear across different magnifications, staining conditions, fields of view, and local tissue structures.
We further construct a standardized WSI-level benchmark to evaluate pathology foundation models under a broad and reproducible setting. The benchmark includes 161 tasks from 44 public datasets and is organized into four coarse-grained clinical domains and nine fine-grained task categories. Compared with single-task or single-split evaluations, the fixed $5 \times 4$ patient-level cross-validation protocol and task-level statistical ranking score provide a more stable basis for model comparison. This design allows performance to be analyzed not only by overall mean performance, but also by task category, clinical domain, and statistically supported task-level ranking.

Across this benchmark, DaX achieves the highest mean performance across tasks and consistently high task-level ranking scores. The results suggest that modeling magnification continuity, local--global tissue correspondence, acquisition robustness, and dense feature geometry improves the transferability of pathology representations. The observed gains are not limited to morphology-dominant diagnostic tasks, but also extend to more challenging endpoints such as biomarker prediction, treatment response, and prognosis, where the relationship between H\&E morphology and labels may be weaker or more indirect.

Several limitations define the scope of the present study. First, the benchmark is based on retrospective public datasets, and prospective validation in clinically locked multi-institutional cohorts remains necessary. Second, the current evaluation focuses on WSI-level prediction and does not fully characterize dense or spatial tasks such as cell segmentation, tissue parsing, registration, detection, or spatial transcriptomics alignment. Third, DaX is evaluated as a visual foundation model and does not incorporate pathology reports, genomic profiles, or structured clinical variables during pretraining. Future work should extend DaX toward multimodal pretraining, expand evaluation to dense and spatial tasks, and perform systematic failure analysis across tissue types, magnifications, staining styles, acquisition conditions, and clinical task categories.

%% file: sec/appendix.tex
\appendix

\input{table/table_1}
\newpage
\input{table/table_2}

%% file: table/table_1.tex
\providecommand{\pfmtabledatasetcell}[1]{\parbox[c]{0.16\textwidth}{\centering\strut #1\strut}}
\par\medskip
\refstepcounter{table}
\label{tab_1}
\begingroup
\normalsize
\noindent\textbf{Table~\thetable. Two-level task taxonomy and complete task inventory of the DaX benchmark.} 
\par\vspace{0.6em}
\endgroup

\begingroup
\makeatletter
\def\@currentlabel{\thetable A}
\label{tab_1a}
\makeatother
\small
\noindent\textbf{A}\quad Hierarchical abbreviation map of the benchmark task taxonomy. The four Level-1 task families correspond to the coarse-grained clinical domains used in the benchmark. These families are further divided into nine Level-2 task categories, which are used to summarize task-level WSI distributions and category-level model performance.
\par\vspace{0.35em}
\setlength{\tabcolsep}{3pt}
\renewcommand{\arraystretch}{1.08}
\setlength{\LTleft}{0pt}
\setlength{\LTright}{0pt}
\begin{longtable}{@{\extracolsep{\fill}}cp{0.27\textwidth}cp{0.43\textwidth}@{}}
\toprule
\makecell{\textbf{Level-1}\\\textbf{Abbr.}} &
\makecell{\textbf{Level-1}\\\textbf{Task Family}} &
\makecell{\textbf{Level-2}\\\textbf{Abbr.}} &
\makecell{\textbf{Level-2}\\\textbf{Task Family}} \\
\midrule
\multirow{3}{0.11\textwidth}{\centering L1-DX} & \multirow{3}{0.27\textwidth}{Diagnostic Pathology} & C01 & Disease Entity \& Histologic Diagnosis \\
 & & C02 & Histologic Grading \& Dysplasia Severity \\
 & & C03 & Anatomic Staging, Invasion \& Metastatic Burden \\
\midrule
\multirow{3}{0.11\textwidth}{\centering L1-BM} & \multirow{3}{0.27\textwidth}{Biomarker \& Molecular Profiling} & C04 & Genomic, Epigenetic \& Pathway Alterations \\
 & & C05 & Single-Marker IHC / Receptor Expression \\
 & & C06 & Composite Molecular \& Immune Phenotypes \\
\midrule
\multirow{1}{0.11\textwidth}{\centering L1-SP} & \multirow{1}{0.27\textwidth}{Tissue / Specimen Context} & C07 & Tissue/Specimen Origin \& Sampling Context \\
\midrule
\multirow{2}{0.11\textwidth}{\centering L1-RP} & \multirow{2}{0.27\textwidth}{Risk, Response \& Prognosis} & C08 & Treatment Response \& Residual Disease \\
 & & C09 & Survival, Recurrence \& Progression Outcomes \\
\bottomrule
\end{longtable}
\endgroup
\addtocounter{table}{-1}

\begingroup
\makeatletter
\def\@currentlabel{\thetable B}
\label{tab_1b}
\makeatother
\small
\noindent\textbf{B}\quad Detailed inventory of the 161 pathology benchmark tasks collected from 44 public datasets. For each task, the table reports the source dataset, task name, prediction type, Level-1 and Level-2 task-family assignments, and the corresponding numbers of patients and WSIs. This task inventory provides the basis for the dataset and task statistics in Figure \ref{fig_1} and for the category-level performance analyses and task-level ranking heatmaps in Figure \ref{fig_3}.
\par\vspace{0.35em}
\setlength{\tabcolsep}{2pt}

\renewcommand{\arraystretch}{1.45}
\setlength{\LTleft}{0pt}
\setlength{\LTright}{0pt}
\begin{longtable}{@{\extracolsep{\fill}}p{0.16\textwidth}p{0.31\textwidth}ccccc@{}}

\toprule
\multicolumn{1}{c}{\textbf{Dataset Name}} & \multicolumn{1}{c}{\textbf{Task Name}} & \textbf{Type} & \makecell{\textbf{Level-1}\\\textbf{Family}} & \makecell{\textbf{Level-2}\\\textbf{Family}} & \makecell{\textbf{Patient}\\\textbf{Number}} & \makecell{\textbf{Slide}\\\textbf{Number}} \\
\midrule
\endfirsthead
\multicolumn{7}{l}{\textbf{B}\quad Detailed inventory of the 161 pathology benchmark tasks collected from 44 public datasets.(continued).}\\
\toprule
\multicolumn{1}{c}{\textbf{Dataset Name}} & \multicolumn{1}{c}{\textbf{Task Name}} & \textbf{Type} & \makecell{\textbf{Level-1}\\\textbf{Family}} & \makecell{\textbf{Level-2}\\\textbf{Family}} & \makecell{\textbf{Patient}\\\textbf{Number}} & \makecell{\textbf{Slide}\\\textbf{Number}} \\
\midrule
\endhead
\midrule
\multicolumn{7}{r}{Continued on next page}\\
\endfoot
\bottomrule
\endlastfoot
\multirow{9}{0.16\textwidth}{\pfmtabledatasetcell{AURORA-MBC~\citep{AURORA-MBC}}} & hla a alteration status & cls-2 & L1-BM & C04 & 46 & 131 \\*
 & hla a hypermethylation status & cls-2 & L1-BM & C04 & 46 & 131 \\*
 & mhc alteration status & cls-2 & L1-BM & C04 & 46 & 131 \\*
 & original receptor subtype & cls-3 & L1-BM & C06 & 45 & 108 \\*
 & pam50 subtype & cls-3 & L1-BM & C06 & 45 & 106 \\*
 & sample collection timepoint & cls-3 & L1-SP & C07 & 49 & 152 \\*
 & sample origin type & cls-2 & L1-SP & C07 & 49 & 152 \\*
 & tapbp hypermethylation status & cls-2 & L1-BM & C04 & 46 & 131 \\*
 & tnbc status & cls-2 & L1-BM & C06 & 46 & 118 \\
\midrule
\multirow{4}{0.16\textwidth}{\pfmtabledatasetcell{BC-Therapy~\citep{BC-Therapy}}} & er status & cls-2 & L1-BM & C05 & 166 & 166 \\*
 & her2 status & cls-2 & L1-BM & C05 & 166 & 166 \\*
 & histologic grade & cls-2 & L1-DX & C02 & 166 & 166 \\*
 & residual cancer burden & cls-4 & L1-RP & C08 & 159 & 159 \\
\midrule
\multirow{7}{0.16\textwidth}{\pfmtabledatasetcell{BCNB~\citep{BCNB}}} & axillary lymph node status & cls-3 & L1-DX & C03 & 1055 & 1055 \\*
 & er status & cls-2 & L1-BM & C05 & 1055 & 1055 \\*
 & her2 expression & cls-4 & L1-BM & C05 & 1055 & 1055 \\*
 & her2 status & cls-2 & L1-BM & C05 & 1055 & 1055 \\*
 & histologic grade & cls-3 & L1-DX & C02 & 924 & 924 \\*
 & molecular subtype & cls-4 & L1-BM & C06 & 1055 & 1055 \\*
 & pr status & cls-2 & L1-BM & C05 & 1055 & 1055 \\
\midrule
\pfmtabledatasetcell{BOEHMK~\citep{BOEHMK}} & progression free survival & survival & L1-RP & C09 & 183 & 183 \\
\midrule
\multirow{2}{0.16\textwidth}{\pfmtabledatasetcell{BRACS~\citep{BRACS}}} & coarse diagnosis & cls-3 & L1-DX & C01 & 189 & 545 \\*
 & fine diagnosis & cls-7 & L1-DX & C01 & 189 & 545 \\
\midrule
\multirow{4}{0.16\textwidth}{\pfmtabledatasetcell{Brain-Mets-Lung-MRI-Path-Segs~\citep{Brain-Mets-Lung-MRI-Path-Segs}}} & egfr mutation status & cls-2 & L1-BM & C04 & 103 & 111 \\*
 & extracranial metastasis at diagnosis & cls-2 & L1-DX & C03 & 103 & 111 \\*
 & gpa histology class & cls-3 & L1-DX & C01 & 103 & 111 \\*
 & pdl1 status & cls-2 & L1-BM & C04 & 103 & 111 \\
\midrule
\multirow{4}{0.16\textwidth}{\pfmtabledatasetcell{CAMELYON16~\citep{CAMELYON16}}} & histology type & cls-3 & L1-DX & C01 & 124 & 124 \\*
 & in situ status & cls-2 & L1-DX & C03 & 125 & 125 \\*
 & lesion size class & cls-3 & L1-DX & C03 & 127 & 127 \\*
 & tumor presence status & cls-2 & L1-DX & C01 & 396 & 396 \\
\midrule
\multirow{2}{0.16\textwidth}{\pfmtabledatasetcell{CAMELYON17~\citep{CAMELYON17}}} & lesion size class & cls-4 & L1-DX & C03 & 200 & 1000 \\*
 & pathologic n stage & cls-5 & L1-DX & C03 & 200 & 1000 \\
\midrule
\multirow{6}{0.16\textwidth}{\pfmtabledatasetcell{CHIMERA2025~\citep{CHIMERA2025}}} & brs risk group & cls-3 & L1-RP & C08 & 131 & 131 \\*
 & lymphovascular invasion & cls-2 & L1-DX & C03 & 131 & 131 \\*
 & progression status & cls-2 & L1-RP & C09 & 131 & 131 \\*
 & repeat tur status & cls-2 & L1-SP & C07 & 129 & 129 \\*
 & tumor substage & cls-2 & L1-DX & C03 & 125 & 125 \\*
 & variant histology status & cls-2 & L1-DX & C01 & 131 & 131 \\
\midrule
\multirow{3}{0.16\textwidth}{\pfmtabledatasetcell{CMB~\citep{CMB}}} & hematologic subtype & cls-2 & L1-DX & C01 & 98 & 290 \\*
 & lung cancer subtype & cls-2 & L1-DX & C01 & 93 & 136 \\*
 & primary site group & cls-5 & L1-SP & C07 & 370 & 771 \\
\midrule
\pfmtabledatasetcell{CPTAC-ALL~\citep{CPTAC}} & organ of origin & cls-10 & L1-SP & C07 & 1061 & 2154 \\
\midrule
\multirow{3}{0.16\textwidth}{\pfmtabledatasetcell{CPTAC-BRCA~\citep{CPTAC-BRCA}}} & immune class & cls-3 & L1-BM & C06 & 103 & 112 \\*
 & pik3ca mutation status & cls-2 & L1-BM & C04 & 103 & 112 \\*
 & tp53 mutation status & cls-2 & L1-BM & C04 & 103 & 112 \\
\midrule
\multirow{5}{0.16\textwidth}{\pfmtabledatasetcell{CPTAC-CCRCC~\citep{CPTAC-CCRCC}}} & bap1 mutation status & cls-2 & L1-BM & C04 & 103 & 245 \\*
 & immune class & cls-3 & L1-BM & C06 & 103 & 245 \\*
 & overall survival & survival & L1-RP & C09 & 94 & 218 \\*
 & pbrm1 mutation status & cls-2 & L1-BM & C04 & 103 & 245 \\*
 & vhl mutation status & cls-2 & L1-BM & C04 & 103 & 245 \\
\midrule
\multirow{9}{0.16\textwidth}{\pfmtabledatasetcell{CPTAC-COAD~\citep{CPTAC-COAD}}} & acvr2a mutation status & cls-2 & L1-BM & C04 & 94 & 98 \\*
 & apc mutation status & cls-2 & L1-BM & C04 & 94 & 98 \\*
 & arid1a mutation status & cls-2 & L1-BM & C04 & 94 & 98 \\*
 & immune class & cls-3 & L1-BM & C06 & 94 & 98 \\*
 & kras mutation status & cls-2 & L1-BM & C04 & 94 & 98 \\*
 & microsatellite instability status & cls-2 & L1-BM & C04 & 93 & 97 \\*
 & pik3ca mutation status & cls-2 & L1-BM & C04 & 94 & 98 \\*
 & setd1b mutation status & cls-2 & L1-BM & C04 & 94 & 98 \\*
 & tp53 mutation status & cls-2 & L1-BM & C04 & 94 & 98 \\
\midrule
\multirow{3}{0.16\textwidth}{\pfmtabledatasetcell{CPTAC-GBM~\citep{CPTAC-GBM}}} & egfr mutation status & cls-2 & L1-BM & C04 & 99 & 243 \\*
 & immune class & cls-3 & L1-BM & C06 & 99 & 243 \\*
 & tp53 mutation status & cls-2 & L1-BM & C04 & 99 & 243 \\
\midrule
\multirow{4}{0.16\textwidth}{\pfmtabledatasetcell{CPTAC-HNSC~\citep{CPTAC-HNSC}}} & casp8 mutation status & cls-2 & L1-BM & C04 & 107 & 256 \\*
 & histologic grade & cls-3 & L1-DX & C02 & 107 & 256 \\*
 & immune class & cls-3 & L1-BM & C06 & 107 & 256 \\*
 & overall survival & survival & L1-RP & C09 & 102 & 243 \\
\midrule
\multirow{4}{0.16\textwidth}{\pfmtabledatasetcell{CPTAC-LSCC~\citep{CPTAC-LSCC}}} & arid1a mutation status & cls-2 & L1-BM & C04 & 108 & 304 \\*
 & histologic grade & cls-2 & L1-DX & C02 & 104 & 292 \\*
 & immune class & cls-3 & L1-BM & C06 & 108 & 304 \\*
 & keap1 mutation status & cls-2 & L1-BM & C04 & 108 & 304 \\
\midrule
\multirow{6}{0.16\textwidth}{\pfmtabledatasetcell{CPTAC-LUAD~\citep{CPTAC-LUAD}}} & egfr mutation status & cls-2 & L1-BM & C04 & 108 & 324 \\*
 & immune class & cls-3 & L1-BM & C06 & 108 & 324 \\*
 & kras mutation status & cls-2 & L1-BM & C04 & 102 & 312 \\*
 & overall survival & survival & L1-RP & C09 & 105 & 313 \\*
 & stk11 mutation status & cls-2 & L1-BM & C04 & 108 & 324 \\*
 & tp53 mutation status & cls-2 & L1-BM & C04 & 108 & 324 \\
\midrule
\pfmtabledatasetcell{CPTAC-LUNG} & lung cancer subtype & cls-2 & L1-DX & C01 & 206 & 604 \\
\midrule
\pfmtabledatasetcell{CPTAC-OV~\citep{CPTAC-OV}} & immune class & cls-3 & L1-BM & C06 & 51 & 160 \\
\midrule
\multirow{3}{0.16\textwidth}{\pfmtabledatasetcell{CPTAC-PDAC~\citep{CPTAC-PDAC}}} & immune class & cls-3 & L1-BM & C06 & 105 & 242 \\*
 & overall survival & survival & L1-RP & C09 & 97 & 227 \\*
 & smad4 mutation status & cls-2 & L1-BM & C04 & 105 & 242 \\
\midrule
\multirow{3}{0.16\textwidth}{\pfmtabledatasetcell{CPTAC-UCEC~\citep{CPTAC-UCEC}}} & ctnnb1 mutation status & cls-2 & L1-BM & C04 & 93 & 93 \\*
 & immune class & cls-3 & L1-BM & C06 & 94 & 94 \\*
 & pten mutation status & cls-2 & L1-BM & C04 & 93 & 93 \\
\midrule
\pfmtabledatasetcell{DHMC-LUAD~\citep{DHMC-LUAD}} & lung adenocarcinoma subtype & cls-5 & L1-DX & C01 & 143 & 143 \\
\midrule
\multirow{5}{0.16\textwidth}{\pfmtabledatasetcell{DLBCL-Morphology~\citep{DLBCL-Morphology}}} & bcl2 fish status & cls-2 & L1-BM & C04 & 114 & 159 \\*
 & bcl6 fish status & cls-2 & L1-BM & C04 & 125 & 173 \\*
 & hans subtype & cls-2 & L1-BM & C06 & 133 & 184 \\*
 & overall survival & survival & L1-RP & C09 & 148 & 202 \\*
 & tumor stage & cls-4 & L1-DX & C03 & 148 & 202 \\
\midrule
\multirow{3}{0.16\textwidth}{\pfmtabledatasetcell{EBRAINS~\citep{EBRAINS}}} & diagnosis & cls-30 & L1-DX & C01 & 2142 & 2312 \\*
 & diagnosis group & cls-12 & L1-DX & C01 & 2142 & 2312 \\*
 & idh status & cls-2 & L1-BM & C04 & 772 & 847 \\
\midrule
\multirow{8}{0.16\textwidth}{\pfmtabledatasetcell{HANCOCK~\citep{HANCOCK}}} & lymphovascular invasion & cls-2 & L1-DX & C03 & 694 & 702 \\*
 & overall survival & survival & L1-RP & C09 & 200 & 200 \\*
 & perineural invasion & cls-2 & L1-DX & C03 & 694 & 702 \\*
 & primary tumor site & cls-4 & L1-SP & C07 & 693 & 701 \\*
 & sample origin type & cls-2 & L1-SP & C07 & 673 & 681 \\*
 & scc keratinizing grade & cls-2 & L1-DX & C02 & 382 & 386 \\*
 & scc nonkeratinizing grade & cls-2 & L1-DX & C02 & 74 & 76 \\*
 & vascular invasion & cls-2 & L1-DX & C03 & 694 & 702 \\
\midrule
\pfmtabledatasetcell{HER2-Tumor-ROIs~\citep{HER2-Tumor-ROIs}} & treatment response & cls-2 & L1-RP & C08 & 85 & 85 \\
\midrule
\multirow{3}{0.16\textwidth}{\pfmtabledatasetcell{Hungarian-CRC-Screen~\citep{Hungarian-CRC-Screen}}} & colorectal cancer status & cls-2 & L1-DX & C01 & 200 & 200 \\*
 & neoplastic status & cls-2 & L1-DX & C01 & 200 & 200 \\*
 & polyp type & cls-3 & L1-DX & C01 & 127 & 127 \\
\midrule
\pfmtabledatasetcell{IMP~\citep{IMP}} & cervical dysplasia grade & cls-3 & L1-DX & C02 & 5320 & 5320 \\
\midrule
\pfmtabledatasetcell{IMP-Cervical~\citep{IMP-Cervical}} & cervical dysplasia grade & cls-4 & L1-DX & C02 & 599 & 599 \\
\midrule
\multirow{5}{0.16\textwidth}{\pfmtabledatasetcell{IPD-Brain~\citep{IPD-Brain}}} & atrx status & cls-2 & L1-BM & C04 & 327 & 483 \\*
 & glioma subtype & cls-3 & L1-DX & C01 & 327 & 483 \\*
 & histologic grade & cls-2 & L1-DX & C02 & 327 & 483 \\*
 & idh1 r132h status & cls-2 & L1-BM & C04 & 327 & 483 \\*
 & p53 status & cls-2 & L1-BM & C04 & 327 & 483 \\
\midrule
\multirow{2}{0.16\textwidth}{\pfmtabledatasetcell{MBC~\citep{MBC}}} & overall survival & survival & L1-RP & C09 & 75 & 96 \\*
 & recist response & cls-4 & L1-RP & C08 & 76 & 97 \\
\midrule
\multirow{3}{0.16\textwidth}{\pfmtabledatasetcell{MUT-HET-RCC~\citep{MUT-HET-RCC}}} & bap1 mutation status & cls-2 & L1-BM & C04 & 1236 & 1236 \\*
 & pbrm1 mutation status & cls-2 & L1-BM & C04 & 1236 & 1236 \\*
 & setd2 mutation status & cls-2 & L1-BM & C04 & 1236 & 1236 \\
\midrule
\pfmtabledatasetcell{NADT-Prostate~\citep{NADT-Prostate}} & treatment response & cls-2 & L1-RP & C08 & 36 & 447 \\
\midrule
\multirow{13}{0.16\textwidth}{\pfmtabledatasetcell{NLST~\citep{NLST}}} & clinical m stage & cls-2 & L1-DX & C03 & 392 & 1065 \\*
 & clinical n stage & cls-3 & L1-DX & C03 & 378 & 1056 \\*
 & clinical stage & cls-4 & L1-DX & C03 & 382 & 1039 \\*
 & clinical substage & cls-7 & L1-DX & C03 & 382 & 1039 \\*
 & clinical t stage & cls-3 & L1-DX & C03 & 403 & 1101 \\*
 & differentiation grade & cls-3 & L1-DX & C02 & 347 & 1002 \\*
 & lung cancer grade & cls-3 & L1-DX & C02 & 308 & 883 \\*
 & pathologic m stage & cls-2 & L1-DX & C03 & 372 & 1067 \\*
 & pathologic n stage & cls-3 & L1-DX & C03 & 357 & 1031 \\*
 & pathologic stage & cls-4 & L1-DX & C03 & 346 & 1008 \\*
 & pathologic substage & cls-7 & L1-DX & C03 & 346 & 1008 \\*
 & pathologic t stage & cls-3 & L1-DX & C03 & 369 & 1063 \\*
 & summary stage & cls-3 & L1-DX & C03 & 96 & 205 \\
\midrule
\pfmtabledatasetcell{OV-Bevacizumab~\citep{OV-Bevacizumab}} & treatment response & cls-2 & L1-RP & C08 & 36 & 85 \\
\midrule
\pfmtabledatasetcell{PANDA~\citep{PANDA}} & isup grade & cls-6 & L1-DX & C02 & 9527 & 9527 \\
\midrule
\pfmtabledatasetcell{POST-NAT-BRCA~\citep{POST-NAT-BRCA}} & lymphovascular invasion & cls-2 & L1-DX & C03 & 50 & 53 \\
\midrule
\multirow{3}{0.16\textwidth}{\pfmtabledatasetcell{PTRC-HGSOC~\citep{PTRC-HGSOC}}} & sample origin type & cls-2 & L1-SP & C07 & 158 & 347 \\*
 & treatment response & cls-2 & L1-RP & C08 & 158 & 347 \\*
 & tumor stage & cls-2 & L1-RP & C08 & 154 & 339 \\
\midrule
\pfmtabledatasetcell{SLN-Breast~\citep{SLN-Breast}} & sample origin type & cls-2 & L1-SP & C07 & 130 & 130 \\
\midrule
\multirow{10}{0.16\textwidth}{\pfmtabledatasetcell{SURGEN~\citep{SURGEN}}} & braf mutation status & cls-2 & L1-BM & C04 & 388 & 388 \\*
 & braf v600e status & cls-2 & L1-BM & C04 & 218 & 218 \\*
 & five year mortality status & cls-2 & L1-RP & C09 & 387 & 387 \\*
 & kras mutation status & cls-2 & L1-BM & C04 & 260 & 260 \\*
 & microsatellite instability status & cls-2 & L1-BM & C04 & 79 & 79 \\*
 & mismatch repair status & cls-2 & L1-BM & C04 & 389 & 389 \\*
 & nras mutation status & cls-2 & L1-BM & C04 & 220 & 220 \\*
 & overall survival & survival & L1-RP & C09 & 144 & 144 \\*
 & ras mutation status & cls-2 & L1-BM & C04 & 389 & 389 \\*
 & tumor stage & cls-2 & L1-DX & C03 & 271 & 271 \\
\midrule
\pfmtabledatasetcell{UCLA-Lung~\citep{UCLA-Lung}} & progression status & cls-2 & L1-RP & C09 & 109 & 109 \\
\midrule
\multirow{9}{0.16\textwidth}{\pfmtabledatasetcell{VAL-BRAF-CRC~\citep{VALBRAFCRC}}} & histologic grade & cls-3 & L1-DX & C02 & 128 & 130 \\*
 & lymphovascular invasion & cls-2 & L1-DX & C03 & 109 & 111 \\*
 & microsatellite instability status & cls-2 & L1-BM & C04 & 134 & 134 \\*
 & mismatch repair status & cls-2 & L1-BM & C04 & 129 & 131 \\*
 & overall survival & survival & L1-RP & C09 & 168 & 168 \\*
 & progression free survival & survival & L1-RP & C09 & 168 & 168 \\*
 & ras braf molecular subtype & cls-2 & L1-BM & C06 & 96 & 96 \\*
 & synaptophysin status & cls-2 & L1-BM & C05 & 129 & 131 \\*
 & tils level & cls-2 & L1-BM & C06 & 129 & 131 \\
\midrule
\multirow{2}{0.16\textwidth}{\pfmtabledatasetcell{VisioMel~\citep{VisioMel}}} & overall relapse status & cls-2 & L1-RP & C09 & 1125 & 1125 \\*
 & relapse without previous melanoma & cls-2 & L1-RP & C09 & 516 & 516 \\
\end{longtable}
\endgroup
\addtocounter{table}{-1}

%% file: table/table_2.tex
\begingroup
\captionsetup{font=normalsize}
\providecommand{\pfmmodelcite}[1]{%
  \if\relax\detokenize{#1}\relax\else~\citep{#1}\fi
}
\providecommand{\pfmmodeltypecell}[1]{\makebox[\linewidth][c]{\normalsize\shortstack[c]{#1}}}
\label{tab_2}
\small
\setlength{\tabcolsep}{1pt}
\renewcommand{\arraystretch}{2.0}
\setlength{\LTleft}{0pt}
\setlength{\LTright}{0pt}
\begin{longtable}{@{}>{\centering\arraybackslash}m{0.090\textwidth}>{\centering\arraybackslash}m{0.180\textwidth}>{\centering\arraybackslash}m{0.105\textwidth}>{\centering\arraybackslash}m{0.120\textwidth}>{\centering\arraybackslash}m{0.075\textwidth}>{\centering\arraybackslash}m{0.120\textwidth}>{\centering\arraybackslash}m{0.060\textwidth}>{\centering\arraybackslash}m{0.215\textwidth}@{}}
\caption{The table summarizes all models included in the benchmark. This table provides the model-level context for the scale--performance analysis in Figure~\ref{fig_3}A and the task-level ranking heatmaps in Figure~\ref{fig_3}D.}\\

\toprule
\shortstack[c]{\textbf{Model}\\\textbf{Type}} &
\textbf{Model Name} &
\shortstack[c]{\textbf{Pretraining}\\\textbf{Method}} &
\shortstack[c]{\textbf{Pretrain}\\\textbf{Mag}} &
\shortstack[c]{\textbf{Pretrain}\\\textbf{Res}} &
\shortstack[c]{\textbf{Patch Model}\\\textbf{Architecture}} &
\textbf{Params.} &
\textbf{Data Statistics} \\
\midrule
\endfirsthead
\caption[]{Evaluated foundation models and their pretraining characteristics (continued).}\\
\toprule
\shortstack[c]{\textbf{Model}\\\textbf{Type}} &
\textbf{Model Name} &
\shortstack[c]{\textbf{Pretraining}\\\textbf{Method}} &
\shortstack[c]{\textbf{Pretrain}\\\textbf{Mag}} &
\shortstack[c]{\textbf{Pretrain}\\\textbf{Res}} &
\shortstack[c]{\textbf{Patch Model}\\\textbf{Architecture}} &
\textbf{Params.} &
\textbf{Data Statistics} \\
\midrule
\endhead
\midrule
\multicolumn{8}{r}{Continued on next page}\\
\endfoot
\bottomrule
\endlastfoot
\multirow{20}{0.090\textwidth}{\pfmmodeltypecell{Pathology\\Vision}} & DaX & \textbf{\normalsize DINOv3} & (2.5,5,10,20)x & 384, 512, 768 & ViT-L/16 & 304M & 104,569 WSIs \\
& DaX-Base & \textbf{\normalsize DINOv3} & (2.5,5,10,20)x & 384, 512, 768 & ViT-B/16 & 86M & 104,569 WSIs \\
 & H-Optimus-1 \citep{hoptimus1} & DINOv2 & 20x & 224 & ViT-G/14 & 1.1B & 1M WSIs \\
 & H-Optimus-0 \citep{hoptimus0} & DINOv2 & 20x & 224 & ViT-G/14 & 1.1B & 0.5M WSIs \\
 & Prov-GigaPath \citep{Prov-GigaPath} & DINOv2 & 20x & 224 & ViT-G/14 & 1.1B & 171,189 WSIs (1.38B patches) \\
 & Virchow2 \citep{Virchow2} & DINOv2 & (5,10,20,40)x & 224 & ViT-H/14 & 632M & 3.1M WSIs \\
 & Virchow \citep{Virchow} & DINOv2 & 20x & 224 & ViT-H/14 & 632M & 1.5M WSIs \\
 & UNI2 \citep{UNI} & DINOv2 & Unknown & 224 & ViT-H/14 & 632M & 350,000 WSIs \\
 & Hibou-L \citep{Hibou} & DINOv2 & Unknown & 224 & ViT-L/14 & 304M & 1.1M WSIs (1.2B patches) \\
 & UNI \citep{UNI} & DINOv2 & 20x & 256, 512 & ViT-L/16 & 304M & 100,000 WSIs \\
 & Phikon-v2 \citep{Phikon-v2} & DINOv2 & 20x & 224 & ViT-L/16 & 304M & 60,000 WSIs \\
 & Hibou-B \citep{Hibou} & DINOv2 & Unknown & 224 & ViT-B/14 & 86M & 1.1M WSIs (512M patches) \\
 & \makecell[c]{EXAONE \\ Path 2.5 \citep{EXAONE-Path} }  & DINO & Unknown & 224 & ViT-B/14 & 86M & 34,795 WSIs \\
 & Phikon \citep{Phikon} & iBOT & 20x & 224 & ViT-B/16 & 86M & 6,093 WSIs \\
 & \makecell[c]{CTransPath \\ (CHIEF) \citep{CHIEF}} & MoCov3 & 10x & 224 & Swin-T/14 & 27.5M & 60,530 WSIs (15.58M patches) \\
 & Lunit ViT-S \citep{Lunit} & DINO & 20x, 40x & 512 & ViT-S/8 & 21M & 20,994 WSIs (19M patches) \\
  & Kaiko ViT-L/14 \citep{kaiko} & DINOv2 & (5,10,20,40)x & 256 & ViT-L/14 & 304M & 29k WSIs \\
  & Kaiko~ViT-B/\normalsize{16}~\citep{kaiko} & DINO & (5,10,20,40)x & 256 & ViT-B/16 & 86M & 29k WSIs \\
 & Kaiko ViT-B/8 \citep{kaiko} & DINO & (5,10,20,40)x & 256 & ViT-B/8 & 86M & 29k WSIs \\
 & Kaiko ViT-S/16 \citep{kaiko} & DINO & (5,10,20,40)x & 256 & ViT-S/16 & 21M & 29k WSIs \\
 & Kaiko ViT-S/8 \citep{kaiko} & DINO & (5,10,20,40)x & 256 & ViT-S/8 & 21M & 29k WSIs \\
\midrule
\multirow{5}{0.090\textwidth}{\pfmmodeltypecell{VLM}} & \makecell[c]{MUSK  \citep{MUSK}} & BEiT-3 & Unknown & 384 & BEiT-3 & 675M & 33,000 WSIs (50M patches) + 1M image-text pairs \\
 & CONCH v1.5 \citep{CONCHv15} & CoCa & 20x & 512 & ViT-L/16 & 306M & UNI initialized + 1.26M pathology image-caption pairs \\
 & CONCH \citep{CONCH} & CoCa & Unknown & 256 & ViT-B/16 & 90M & 1.17M human pathology image-caption pairs \\
 & PLIP \citep{PLIP} & CLIP & Unknown & 224 & ViT-B/32 & 87M & 208,414 pathology image-text pairs \\
 & BiomedCLIP \citep{BiomedCLIP} & CLIP & Unknown & 224 & ViT-B/16 & 86M & 0.96M radiology, 0.38M pathology \\
\midrule
\pfmmodeltypecell{Expert\\Distillation} & GPFM  \citep{GPFM} & - & Unknown & 224 & ViT-L/14 & 307M & 72,280 WSIs (190M patches) \\
\midrule
\pfmmodeltypecell{Multimodal} & mSTAR  \citep{mSTAR} & - & 20x & 224 & ViT-L/16 & 303M & 11,727 WSIs + 26,169 modality pairs \\
\midrule
\pfmmodeltypecell{ST\\Pathology} & OmiCLIP \citep{OmiCLIP} & CLIP & Unknown & 224 & ViT-L/14 & 304M & 2.18M pathology image--spatial transcriptomics pairs \\
\midrule
\multirow{5}{0.090\textwidth}{\pfmmodeltypecell{Natural\\Image}} & DINOv3 ViT-L \citep{Dinov3} & DINOv3 distillation & - & 224 & ViT-L/16 & 300M & 1,689M \\
 & DINOv3 ViT-B \citep{Dinov3} & DINOv3 distillation & - & 224 & ViT-B/16 & 86M & 1,689M \\
 & DINOv2 ViT-B \citep{Dinov2} & DINOv2 & - & 224, 518 & ViT-B/14 & 86M & 142M \\
 & DINO ViT-B \citep{DINO} & DINO & - & 224 & ViT-B/16 & 85M & 1.28M \\
 & ResNet-50  \citep{ResNet} & - & - & 224 & ResNet-50 & 25.6M & 1.28M \\
\end{longtable}
\endgroup